\documentclass[
 amsmath,amssymb,
 aps,
 pre,floatfix
]{revtex4-2}

\usepackage{float}
\usepackage[utf8]{inputenc}
\usepackage[T1]{fontenc}
\usepackage{amsthm}
\newtheorem{theorem}{Auxiliary result}
\newtheorem{corollary}{Main result}
\newtheorem{definition}{Definition}

\usepackage{bm}
\usepackage{dsfont}
\usepackage{amsfonts}
\usepackage{indentfirst}
\usepackage{graphicx}
\usepackage{dcolumn}
\usepackage{bm}
\usepackage{color}
\usepackage{wasysym}

\begin{document}

\title{Evolution of behaviors in heterogeneous traffic models as driven annealed disorders and its relation to the n-vector model}

\author{Ricardo Sim\~ao}%
 \email{summernightdream@fis.grad.ufmg.br}
 
 \affiliation{Departamento de F\'\i sica, Universidade Federal de Minas Gerais,
Caixa Postal 702, CEP 30161-970, Belo Horizonte - MG, Brazil.
}%

\date{\today}
  
\begin{abstract}
In one-dimensional, heterogeneous systems, the whole traffic dynamics depend strongly on the behavior of the leading vehicle. This result holds for a class of vehicular traffic models satisfying the following properties. The interactions are unidirectional. The dynamics of the particles maximize the velocity or reduces the gap between particles. The particles are hard. We use this result to show a link between traffic models and graphs theory with the consequence that as driving styles spread through social contagion and appear randomly, the inhomogeneities of the system becomes dynamical, or \textit{annealed}, toward specific regions in the space of parameters. Interpreting parameters as entries of vectors defined in the parameters space appear analogies between the evolutionary dynamics of these systems and asymptotic behaviors of the \textit {n-vector model}.  When the time-scale ratio of imitation to the mutation processes, $ \tau_i / \tau_m $, is small an organized state where ``orientation'' corresponds to the set of parameters of the slowest strategies is favored, and if this ratio is big an unorganized state without a preferential orientation is favored. 
\end{abstract}


\maketitle

\section{Introduction}

Vehicular traffic severely impacts our daily lives. Long commute times have implications for the driver's well-being \cite{longcommute,peters2004exposure} and have noticeable economic impacts \cite{weisbrod2003measuring}. Practical solutions in traffic-related problems are often based on insights emerging from understanding the underlying mechanisms of simplified theories. 
 
Since the seminal work of Lighthill and Whitham on traffic flow modeling \cite{lighthill1955kinematicI,lighthill1955kinematic}, much effort has been spent to formulate a comprehensive and concise vehicular traffic theory \cite{kerner2012physics,piccoli2009vehicular,nagatani2002physics,chowdhury2000statistical,kerner1996experimental,maerivoet2005cellular,nagel1996particle,hoogendoorn2001state}. The models can be separated in three categories according to the level of detail\cite{hoogendoorn2001state}: macroscopic models, as the flow model proposed by Lighthill and Whitham; mesoscopic models such as car-following models \cite{brackstone1999car} and  kinetic models \cite{prigogine1971kinetic}; and microscopic models, like cellular automaton (CA) models \cite{maerivoet2005cellular,maerivoet2004non,nagel1996particle,kerner2012physics}. In traffic flow models, one applies concepts from hydrodynamics to traffic considered as a continuous flow. The typical quantities of interest are average velocities, density, and current. In the mesoscopic models, there is a more detailed description of the individual behavior, and some variables are treated as continuous. Microscopic models present the most detailed information on individual behaviors. In CA models, one uses cells, either occupied or empty, to represent the environment. The system vehicles plus drivers are particles that hop cells according to a given set of rules for each individual, which depend on the cell state, the neighbors of that cell, external parameters, and may depend on random factors \cite{wolfram1983statistical,kerner2012physics}.

Papageorgiou was the first to question whether continuous models could reach the quantitative and qualitative accuracy verified in other fields of physics \cite{papageorgiou1998some}. A step toward realism and accuracy in microscopic models is given with the introduction of heterogeneous populations as a homogeneous algorithm submitted to a quenched disorder on its parameters \cite{ben1994kinetics,krug1996phase,helbing2008power,krug2000phase,barma2006driven,ramana2020traffic,ramana2021power}. This variation of parameters brings into discussion the differences in behaviors of the human drivers and vehicles performances of real traffic.

The experimental measurements of the behavior in different places and observation of highly heterogeneous traffic where many types of vehicles coexist suggest that these behaviors are not trivially related by a change in the parameters of a single model. This leads us to introduce multi-behavioral populations as an extension of the concept of heterogeneous populations but also regarding differences in algorithmic nuclei. The particles in these populations have different parameters and may follow different (but environmentally compatible) algorithms. We shall dedicate a section to its analysis and define them in the same way as heterogeneous populations. 

 A specially important instance where the study of heterogeneous populations helps to understand traffic-related phenomena is the emergence of social dilemmas analysis. Many interesting attempts to take social preferences into discussion are made when an additional fast highway is built to connect two previously unconnected locations \cite{hagstrom2001characterizing},taking the fast lane to reduce the time travel, \cite{karlin2017game}, lane changing \cite{iwamura2018complex,tanimoto2014dangerous}, driving lane selection protocols \cite{tanimoto2019improvement}, route selection \cite{tanimoto2016social}, nearby junction-like traffic structures \cite{nakata2010dilemma, yamauchi2009dilemma} and overtaking \cite{simao2021social}. These works show that individual preferences may significantly impact the averaged properties of the system. 

Human drivers are capable and adaptable.  When in social dilemmas, they would change their behaviors to comply with the optimum strategy. As far as we know, the study of the dynamics of the driver's preferences is absent in the literature. Notice that heterogeneous populations cannot simulate this new situation which may be seen as the advent of new drivers, new vehicles with different characteristics, depreciation of the vehicles and the street conditions, and so on because its heterogeneities are static. As a  first approximation to attack this problem, we postulate analogous to the well-known local imitation and the mutation processes but applied to traffic systems, which we shall define precisely in the main text.  Here we justify such similarities as follows. The behavior of the driver is strongly dependent on the behavior of the drivers nearby. Therefore, we model the local imitation dynamics as the transformation of the driver's behavior to the behavior of a better-fit neighbor. In other words, the imitation process is analogous to a death-birth process where a low fitness particle ``dies'' given place to the offspring to a better-fitted neighbor \cite{sigmund1999evolutionary,weibull1997evolutionary,hofbauer1998evolutionary,vincent2005evolutionary,nowak2006evolutionary,perc2013evolutionary,hofbauer2003evolutionary}. The mutation dynamics is related to unpredictable changes in random factors that are not in direct control of the driver which is closely related to the mutation processes in evolutionary theory applied on biology \cite{sigmund1999evolutionary,weibull1997evolutionary,hofbauer1998evolutionary,vincent2005evolutionary,nowak2006evolutionary,ferreira2002mutation,foster1990stochastic}. Notice that the evolutionary process of behaviors presented here is equivalent to the introduction of heterogeneity by \textit{annealed disorders} on the parameters but \textit{driven} towards behaviors (or set of parameters) the evolution favors.

One may point out that if the distribution of characteristics of the vehicles is similar, then cultural preferences and law enforcement are important to the evolution of the driver's behaviors in different places. For example, in single-laned streets is socially acceptable to use all the space available to overtake in India, but the same behavior is punishable in Germany. In the former, the evolution of behaviors is driven by \textit{reward}, but it is driven by \textit{punishment} in the latter.  We are now in a position to formulate our main question as follows. Suppose we have a heterogeneous model able to simulate the average behavior of drivers using a number $m$ of external parameters within a given error margin, where does an arbitrary evolutionary process leads this population?

All previously cited works regarding heterogeneous populations present different algorithms, but a recurring feature is the spontaneous formation of clusters behind slow leaders. In the survey of these works, we synthesized three simple properties they all share and heuristically show that any uni-dimensional model satisfying those properties should behave the same way. Surprisingly, this result simplifies considerably the evolution of behaviors on very general grounds in such a way that the evolutionary dynamics no longer depend on the hidden assumptions fomenting it: it is the same whether it is rewarding-based or punishing-based. The stable cluster formation in heterogeneous populations implies a link to graph theory that rewards us with a link to a well-known problem in theoretical evolution: the evolution of a population spatially disposed on a directed-linear graph \cite{lieberman2005evolutionary,szabo2007evolutionary,shakarian2012review,fu2009evolutionary,lieberman2005evolutionary,nowak2004evolutionary}. The leader always wins the evolutionary struggle.

Regarding the parameters of a given particle as entries of a vector offers a parallel between evolutionary dynamics and statistical mechanics. The spontaneous cluster formation result and the dynamical effects of the evolutionary processes imply analogies between the evolutionary dynamics of this system and the temporal evolution of the \textit{n-vector} model, where $n=m$ is the number of parameters suffering variation in the model. Indeed, there is either an orientation ordering of these ``vectors'' or randomization of the orientations when imitation or mutation are, respectively, the dominating processes. We shall present this discussion in more detail in the main text, including the reason why the similarity is closer to the \textit{n-vector model} than to active-matter models such as the presented by Vicseck and Chatè \cite{vicsek2012collective,gregoire2003moving,doostmohammadi2018active,gregoire2004onset}. These results are also valid to multi-behavioral populations.

We organize this paper as follows. In section \ref{GenRes} we state the premises that determine a group of models we are working with and derive results regarding the equilibrium attained in heterogeneous populations that we shall use to obtain the two main results of this work. We just enunciate the results, the arguments are heuristic and highly abstract, so we present them in the \textbf{Appendix}  and give two concrete examples in the main text.  We present the first in section \ref{ex1}. We use the heterogeneous version of the NaSch model \cite{nagel1992cellular} due to the simplicity of calculations. The simulation results of this model corroborate to the main results and we show a transition from a organized state at $\tau_i/\tau_m \rightarrow 0$ to a random orientation state when $\tau_i/\tau_m $ increase in value, with special attention to the neighborhood of $\tau_i/\tau_m = 1$. We reserve section \ref{ex2} to introduce multi-behavioral populations and verify the results in these models. Unexpected results related to these populations are also discussed in that section. We discuss the results in section \ref{diss}. 


\section{General results}\label{GenRes}

A \textit{particle} an object representing the combination of the vehicle and its driver to us. It occupies $d$ sites and has a set of $m$ intrinsic parameters that control its behavior. The state of the system at time $t$ is defined by the position and the velocity of each particle. The \textit{interaction} between particles takes place when one particle changes its motion in response to another particle. This defines the interaction radius, $R$, as a maximum distance among particles where interactions may occur. We define the interaction radius of the $j$-th particle as its maximum velocity, $R_{j}= v_{max,j}$. In general, let $\chi_{j,i}(t)$ be the distance between the $j$-th and the $i$-th particles at the instant $t$ ($x_j(t) > x_i(t)$), $\chi_{j,i}(t)=x_{j}(t) - x_{i}(t) - d_j \ge 0$, we shall say that if $\chi_{j,i}(t)>R_{j}$ the particles are dynamically independent at the instant $t$. 

We shall consider one-dimensional vehicular traffic models satisfying the following three properties:

\begin{description} 
\item[P1] (Directed interactions) A particle can only affect the dynamics of the particles behind it. 

\item[P2] (Hard-particle) Two particles cannot occupy the same site or pass through each other. 

\item[P3] (Velocity maximization) The dynamics of each particle drive it to either to increases its velocity to a given limit, $v_{max}$, or to reduce the gap to the next particle, observing the properties above.
\end{description}

Property \textbf{P1} is a formulation of the observation that the behavior of the vehicles in front of a driver is much more important to its behavior than the behavior of drivers behind him. Property \textbf{P2} we state that accidents and overtaking are neglected (forbidden). Property \textbf{P3} states that particles strive to have the maximum speed allowed when not interacting (in the free state) and decrease the gap up to a given point to the next particle when interactions are important (in the congested state). This property is a formulation of the daily experience that drivers strive for minimizing the commuting time in a safe way. It worth mentioning that these properties do not fix any algorithm.

Those properties and definitions are compatible with all uni-dimensional CA to the best of our knowledge. We will define some parameters for the sake of clarity.

\begin{definition}
The slowest strategy is the one that reaches the slowest average speed in the given environment.
\end{definition}

\begin{definition}
Let $x_j(t)$ be the position of the $j$-th particle at the time $t$, the measured velocity of $j-$th particle in a time interval $\Delta t=t_2-t_1$ is given by 
\[
v_j=\frac{\Delta x_j}{\Delta t} = \frac{x_j(t_2)-x_j(t_1)}{t_2-t_1}.
\]
\end{definition}

\begin{definition}
 Let $\bar{a}$ be the time-average of the variable $a$. We define $\chi_{j}^{*}$ and $v^{*}$ as the limits $\bar{\chi}_{j} \rightarrow \chi_{j}^{*}$ and $\bar{v}_{j}\rightarrow v^{*}$, for all $1\le j \le N$.  These variables, when the limits exists, characterize a \textbf{steady state} of the system. 
\end{definition}

We shall use those definitions in systems obeying the cited properties. Under periodic boundary conditions and random initial conditions, we can show the following results. 

\begin{theorem}
 In uni-dimensional, heterogeneous, particle-flow systems satisfying the propositions P1, P2, and P3, there is a stable steady state characterized by $\chi_{j}^{*}$  and $v^*$, for all $ 1 \le j \le N $, at densities where a \textbf{leader} can be distinguished. 
\end{theorem}

This result is compatible with real-traffic measurements and is derived in the case of the heterogeneous TASEP model in \cite{evans1997exact}. 

This is a simple but powerful result. In heterogeneous models obeying the premises, exists a stable steady state, where the $j$-th particle is, on average, at a distance $\chi^{*}_{j, i}$ from a following $i$-th particle, and they share the same average velocity. These results will enormously facilitate the understanding of the evolutionary dynamics of behaviors in those systems. We shall disassembly this result in two.

\begin{theorem} (\textit{Connection to Graph Theory}) In systems satisfying the conditions of the \textbf{Auxiliary result 1}, there is a graph, $\Omega$, that may represent the interaction nets on the system. $\Omega$ is a directed, random graph with a single leader.
\end{theorem}

 \begin{figure}[t!]
    \centering
    \includegraphics[scale=.8]{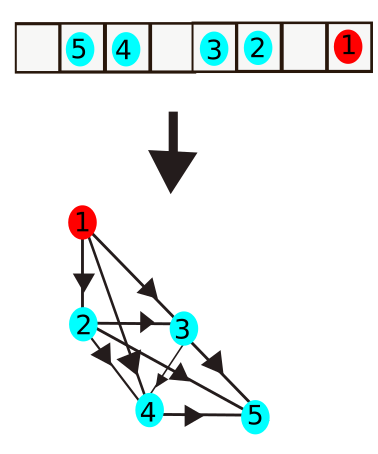}    
    \caption{Graph representation of a simple example. A representation of a possible simulational situation of some model obeying the premises (above). The blue disks are fast-strategies particles and the red disks are slow-strategies. All are numbered with reference to the first and the interaction radius of all is equal to $3$ sites. Below, a representation of the interaction network of the particles in this system. The interactions are directed, as represented by the arrows in the figure, and are not first-neighbors. The connectivity is not shown to simplify the visualization. Notice that the $5$-th particle does not have a link to the leading particle as it is beyond its interaction radius (there are $3$ vehicles between the $5$-th vehicle and the leader so that $\omega_{1,5}=0$ for all $t$) but there is a closed path from the leader to this particle so that the behavior of the leader can be transmitted down this net to the $5$-th particle.}
    \label{Graph}
\end{figure}

 This result is exemplified in figure \ref{Graph}. As $\omega_{i,k} = 0$ for all $t$ and for all $i$ such that $(i-k)>R_k$ in an enumeration where the $N$-th particle is the leader,  then $\omega_{i,k} = 0$ and these particles cannot be linked. If the $i$-th particle is the leader, the reason why  $\omega_{i,k} = 0$ is not geometry, but a consequence of the isolation of the leader. 

One may then point out that $\Omega$ exists even at high densities.  Nevertheless, only under the conditions satisfying the \textbf{Auxiliary result 1} the single-leader and directed properties of these graphs, which will be important in what follows, appear. I this case, if there are many slow particles in the system, we may even divide $\Omega$ in a set of independent graphs, each obeying the properties states by the \textbf{Auxiliary result 2}.

 As the variable used as \textit{payoff} in the works attacking the social dilemma in traffic problem is the average velocity, it is convenient to state here a result regarding this variable.

\begin{theorem}(Maximum velocity limit) In systems satisfying the conditions of the \textbf{Auxiliary result 1}, the average velocity in the steady-state of any particle in a heterogeneous population is not higher than the average velocity of the slowest particle in this population.
\end{theorem}

Assuming that the average velocity is a function of the parameters of the model and an evolutionary dynamics acts to \textit{increase} this variable we can understand the changes of the behavior of the particles as a \textit{drive} of the system to modify the disorders in a way to increase velocity. The system would erase the quenched disorders and stabilize in the set of parameters characterizing this maximum. The auxiliary results imply that this evolutionary development is not as straightforward as it seems.   

\begin{definition}
A local imitation process is any process where the behavior of a randomly chosen particle is transmitted to a neighbor it is interacting over a characteristic interval $\tau_i$. The success probability increases with the target's payoff and it is not null when the payoffs are the same. 
\end{definition}

The details of the process may depend on various complicated sub-processes. It is important to emphasize that definition $5$ also does not fixate the local imitation algorithm. It just states the only happens among interacting particles, its probability increases with the advantage in doing so, and random drift is possible. 

We enunciate the first main result of this work as follows.

\begin{corollary}
(Domination of the slowest) In systems under the conditions of the \textbf{Auxiliary result 1} under a local imitation process, the outcome is a population using the strategy that returns the slowest average velocity present in the initial configuration.
\end{corollary}

This means that if a primordial population of particles following similar strategies were put in a one-dimensional environment, the final evolutionary outcome would be the same, even though different social pressures are in play. Also, in all of the heuristics presented so far was assumed that there is only one slow strategy in the system. It may happen that more slow strategies have the same average velocity in the given conditions. In this case, the results do not rule out coexistence and will be important in the discussion of multi-behavioral populations later. 
\begin{figure}[t!]
    \centering
    \includegraphics[scale=0.4]{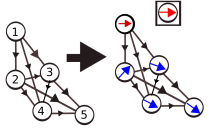}
    \caption{The same as presented in fig \ref{Graph}.  To each node of $\Omega$ we associate a $m$-dimensional vector defined in the space of parameters. We do not present the connectivity of the vectors for simplicity. The leader has the orientation of the slower particle in the population, which is represented by the red arrow in the box. The imitation dynamics transform the orientation of the $2$-th particle aligns with the orientation of the leader in a single event.}
    \label{laticce-graph}
\end{figure}

The uni-dimensional character of the problem and random drift leads the system to its evolutionary fixed points that correspond to the slower velocities in the group.

Let the parameters of a particle be $(x_1,x_2,...x_m)$. We interpret it as the components of a vector in the $m$-dimensional space of parameters. The \textbf{Main result 1} describes a mechanism that drives these vectors to a preferable vector, $(x_1^*,x_2^*,...x_m^*)$. We place the vectors corresponding to the parameters of the particles in the nodes of the graph $\Omega$. The heterogeneity in the system is now an annealed disorder driven to the orientation of the leader, which is parallel to $(x_1^*,x_2^*,...x_m^*)$, as illustrated in figure \ref{laticce-graph}.

  It is possible to trace a path from the leader to the last particle passing through all other particles. Therefore, the evolution of this system closely resembles the evolution of a uni-dimensional \textit{m-vector} model with free boundaries and the first vector (leader) fixed in a given direction. The differences are: the number of interacting neighbors depends on the interaction radius; the interactions are directed, and; the number of interacting particles is not fixed, because as the system evolves, the graph loses its leaders. The first and last differences are not detrimental, as it just replaces a spin chain for another smaller chain with the same properties. The uni-directional character of the interactions also seems not to be detrimental once the orientation of the first vector is fixed and the system does not have ``thermal'' fluctuations.

Let us define mutation. 
\begin{definition}
A mutation process is the independent random change of one of the parameters of the particle to a neighboring value in the space of parameters, occurring at an interval $\tau_m$. 
\end{definition}

The drivers may change their behavior independently of the influence of their neighbors. It can be related to automobile fleet renewal, upgrade the vehicle equipment, reactions to the behavior of other drivers or the road conditions, etc. We enunciate the second main result of this work as follows:

\begin{corollary}
(Domination of the slowest of all ) Systems under the conditions of the \textbf{Auxiliary result 1}, subjected to mutation associated to local imitation dynemics presents two distinct behaviors in the limits $\tau_i/\tau_m \rightarrow \infty$ and $\tau_i/\tau_m \rightarrow 0$.
In the former limit,  the long-time distribution will be analogous to a random walk. In the latter, the slowest strategies reachable in the parameter space will dominate the population.
\end{corollary}
\begin{figure}[t!]
    \centering
    \includegraphics[scale=0.4]{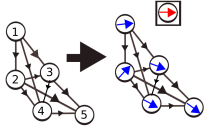}
    \caption{ The same as figure \ref{laticce-graph} but the addition of mutation. Here $\tau_m\gg\tau_i$, and the leader has the closest orientation to the slowest of all strategies in the parameters space that is represented as a red arrow in the box (in this example this orientation is unique, although this is not necessary in the general case). The imitation dynamics align the orientation of the $2$-nd particle to the orientation of the leader in a single event and the mutation dynamics causes a random, small flip around it. In case a big flip occurs such that the node vector becomes closer to the red vector than the leader, it becomes the new leader after the system relaxes in the new configuration.}
    \label{laticce-graph-II}
\end{figure}

The evolution through imitation dynamics selects the better fitted. Here, we see that instead of improving the payoff of the individual, it depletes it. Including mutation,  the evolutionary pressures can lead to the lower payoff possible. As far as we know, these are the only systems these results appear. 

With regard to the comparison to the \textit{m-vector} model,  the mutation tends to randomize the orientations by definition. It seems to have a similar effect as a ``thermal'' noise, but we should be careful here because the leader of the graph $\Omega$ is no longer stationary. If $(x_1^*,x_2^*,...x_m^*)$ is related to the slower strategy possible, then when a fluctuation happens that orient a particle closer to it than the leading particle, the structure of $\Omega$ breaks and reappear with this particle as the leader. If $\tau_m\gg \tau_i$ such that the system can ``thermalize'' in the new configuration, we can suppose $\Omega$ static but with the first vector subjected to an uneven noise that \textit{drives} it to $(x_1^*,x_2^*,...x_m^*)$ as represented in figure \ref{laticce-graph-II}. On the other hand, if $\tau_m\gg \tau_i$ is not true, then the fragmentation of the main cluster happens. 

Interestingly, at high concentrations, $\Omega$ is a cyclical graph and cannot be fragmented. As \textbf{all} particles are connected, the evolution of this system is now more closely related to the \textit{m-vector} model with periodic boundary conditions, and the effects of the mutation dynamics are now identical to thermal noise. We do not have a drive to the orientation of the leader any longer as leaders are nonexistent, but by similarity with to the \textit{m-vector} model, we expect order when $\tau_i/\tau_m \rightarrow 0$ and disorder when $\tau_i/\tau_m \rightarrow \infty$. The nature of the transition from the disorder at $\tau_i/\tau_m \rightarrow \infty$ to order at $\tau_i/\tau_m \rightarrow 0$ at low densities as well as the possible change of behavior of the evolutionary dynamics between low densities and high densities should be addressed and further discussed in future works. 

We end this very abstract discussion we justify the analogy to the \textit{m-vector} model instead to active-matter models \cite{vicsek2012collective,gregoire2003moving,doostmohammadi2018active,gregoire2004onset} that one would though to be fitter as the particles are Mobile. The reason is that very little of the discussion depends on the dynamics of the positions of the particles. Since the \textbf{Auxiliary result 2}, what was important was whether the position of the particle was inside the interaction radius of the particles following it or not, not the real positions. With this fact in mind and interpreting whether or not the particles are inside of the interaction radius from one another as a probability we could associate the dynamical problem to a static problem.  



\section{Example I: heterogeneous NaSch model}\label{ex1}
\begin{figure}[htp!]
    \centering
    \includegraphics[scale=.5]{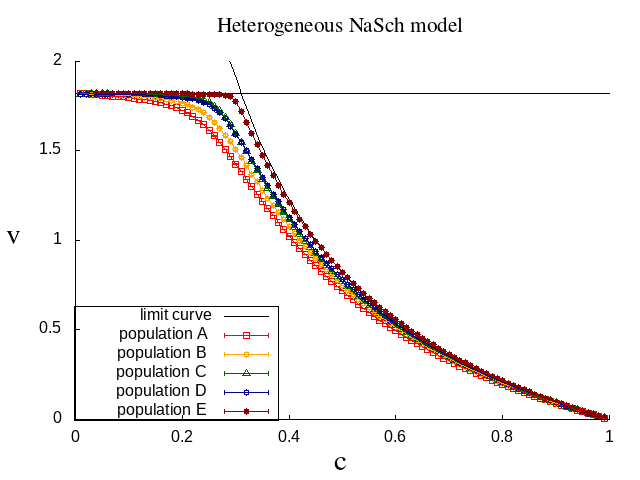}    
    \includegraphics[scale=.5]{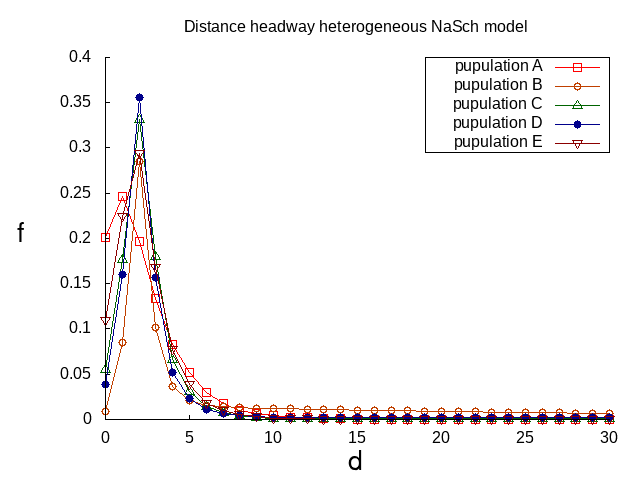}
    \caption{The average velocity of the population as a function of the concentration (above) for mixed populations measured over $10$ initial random configurations. The error bars are smaller than the size of the symbols. The partial concentrations and strategies characterizing each population are presented in the main text. The populations $A$ are a homogeneous slow population. The mean velocities of the heterogeneous populations are localized in the region limited below by the homogeneous slow population curve and limited above by the curves $f(c)=(v_{s}-p_{s})$ and $g(c)=(1-p_s)\cdot(1/c-1)$ represented as solid lines. Notice that the higher the concentration of faster strategies the closer the population is to the superior limit. Below, the frequency histogram of the relative distance between the particles. Notice the peaks at $d=v_{max,s}=2$.}
    \label{HeteroNaSch}
\end{figure}

\begin{figure*}[htp!]
    \centering
    \includegraphics[width=\linewidth]{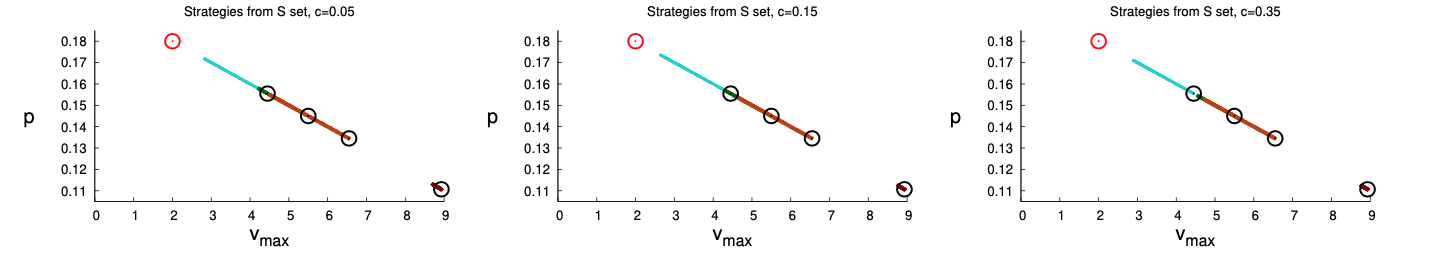}
    \caption{Study of the evolution of the system dictated by the imitation dynamics presented in the text. Panels a), b), and c) show the measurements of the average maximum speed ($ v $) and the average deceleration parameter ($ p $). The black dots surrounded by a circle indicate the initial values of the populations. The red dot surrounded by a red circle is the slower strategy point, $ (2,0.18) $. The colored dots correspond to the averages measured consecutively. Notice the steady path towards the slower strategy. The evolutionary path as a straight line is atypical. Since as the $ (2,0.18) $ strategy expands the average changes in proportional steps in the $ v_ {max} $ ax and the $ p $ ax (see how we build the set \textrm{S}). Panel a) shows the relative data for a very low concentration $ c = 0.05 $, panel b) shows the data for a relatively low concentration $ c = 0.15 $ and panel c) shows the data for a moderate concentration $ c = $ 0.35. The slow convergence of the E population is due to the few slow specimens in this population (if $n_s$ is the number of slow strategies out of $N$ individuals, then the probability for a particle to be selected to the imitation tryout with a slow particle as a target is proportional to $n_{s}/N$).}
    \label{ImitationFixedPoint}
\end{figure*} 
\begin{figure*}[htp!]
    \centering
    \includegraphics[scale=0.33]{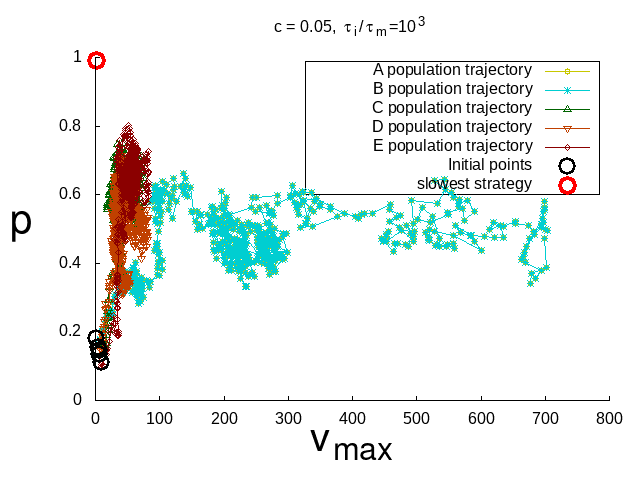}
    \includegraphics[scale=0.33]{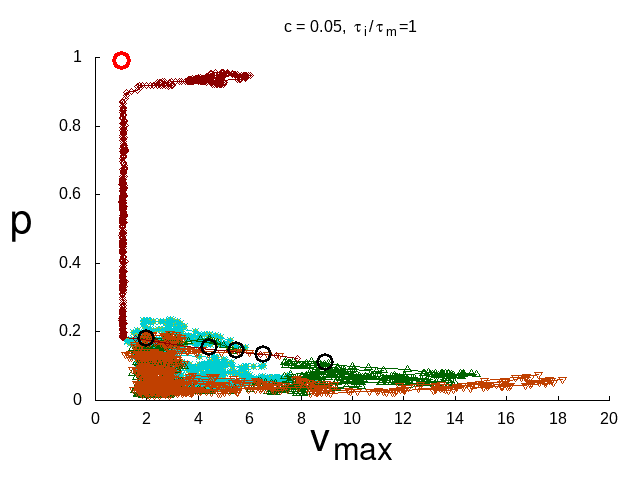}
    \includegraphics[scale=0.33]{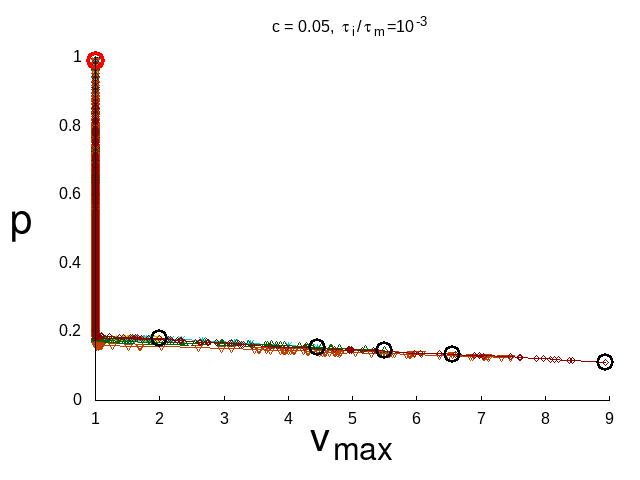}
    \caption{Study of the evolution of the system under the imitation-mutation processes on similar initial conditions as presented in figure \ref{ImitationFixedPoint} at three different values of $\tau_i/\tau_m$ at small concentrations, $c=0.05$. The points are averages of $p$ and $v_{max}$ over the population. The mutation algorithm allows a randomly chosen particle to change either its velocity or deceleration factor to a close value, but the changes are limited to $v_{max} \in [1,\infty)$ and $p \in [0, 0.99]$. The slower possible strategy is $\omega=(1,0.99)$ represented as a red circle. The black circles are the initial points of the populations, the same as presented in figure \ref{ImitationFixedPoint}. Panel a) show the data relative to a situation where the mutation is much more frequent than imitation $\tau_i/\tau_m=10^{3}$. The dynamics are a random walk bounded in $[0.0,.99]$ in the $p$ ax but unbounded on the right in the $v_{max}$ ax. The panel b) show the data relative mutation as frequent as imitation $\tau_i/\tau_m=1$. The velocities do not grow to as high values as in panel a), and the observed behavior is very complicated. In panel c) we show the data relative to mutation much less frequent than imitation $\tau_i/\tau_m=10^{-3}$. In this case, the attraction to the point $(1,0.99)$ is clear. The different time scales of attraction in the parameters $v_{max}$ and $p$, evidenced by the path's almost straight angle and the spacing among the points, can be explained based on equations \ref{media1}. It states that the velocity of the $j$-th particle does not depend strongly on $p_j$. For it to become important, the $j$-th must become the leader. In other words, when every particle has the same $v_{max}$, the particle with higher $p$ leads, but for the system to thermalize in this new configuration takes time, causing the differences in time scales.}
    \label{MutationFixedPoint}
\end{figure*}

In the model proposed by Nagel and Schreckenberg \cite{nagel1992cellular} (NaSch), the particles first accelerate one unity up to the maximum velocity $v(t)=MIN(v(t-1)+1, v_{max})$ . Evaluate the space and in the second step $v(t)=MIN(v(t),\chi(t))$. The is a randomization of the velocities in the third step: $v(t)=MAX(0,v(t)-1)$ with probability $p$. After all particles have their velocity updated, the positions are updated in the fourth step: $\textbf{x(t)}=\textbf{x(t-1)} + \textbf{v(t)}$ (the boldface represents vector notation).

The property \textbf{P3} is observed in the first NaSch step and the properties \textbf{P1} and \textbf{P2} in the second step. See \cite{nagatani2002physics,chowdhury2000statistical,kerner1996experimental,maerivoet2005cellular,nagel1996particle,hoogendoorn2001state} for a thoroughly discussion.

Using the boundary conditions as $x(t)_{N+i}=x(t)_{i}+L$, and the approximation $\bar{v}= (1-p)v + p(v-1)$, we make an approximation for the average velocity in the function of the concentration that works well at low and high concentrations using the overall velocity $\mathcal{V}=\sum_{i=1}^{N}v_i$, to estimate $v$:           

\begin{equation}\label{media1}
    \bar{v}=\frac{\mathcal{V}}{N}=\begin{cases}
    v_{max, s} - p_s & \text{in free state}\\
    (1-p_s)(1/c - 1) & \text{in congested state}.
    \end{cases}
\end{equation} 
This approximation is exact when $p_j=0 \ \forall \ i$. At low densities, where the first step dominates the dynamics, the result is trivial considering that the slow population, $(v_{max,s},p_s)$, leads. In the congested phase, we may neglect the second term in $\bar{v}$ because $(v-1)$ is either zero or non-applicable for the majority of the particles to obtain the result. As the second NaSch step is the more important in all densities in a heterogeneous population, we may approximate the average distance between particles as: 

\begin{equation}\label{media2}
    \chi_{j,j+1}= p_{j}q_{j+1}(\bar{v}_s-1) + p_{j}p_{j+1}\bar{v}_s+ q_{j}p_{j+1}(\bar{v}_s+1).
\end{equation}


Equations \ref{media1} and \ref{media2} are equivalent to the statements of the \textbf{Auxiliary results}. 

As an example, we will consider the following set of strategies:
\[
S=\{(1+n,0.19-0.01n): n \in\{1,2,3,4,5,6,7,8\}\},
\]

The initial distribution of the types in the population is described by the vector $(c_1,c_2,c_3,c_4,c_5,c_6,c_7,c_8)$, where $c_i$ is the initial fraction of type $i$ ($\sum_i c_{i}=1$). We work with five representative initial conditions:
\begin{eqnarray}
A&=&(1, 0, 0, 0, 0, 0, 0, 0) \nonumber \\ 
B&=&(0.50, 0.05, 0.05, 0.05, 0.05, 0.05, 0.05, 0.20) \nonumber \\
C&=&(0.125, 0.125, 0.125, 0.125, 0.125, 0.125, 0.125, 0.125) \nonumber \\
D&=&(0.20, 0.05, 0.05, 0.05, 0.05, 0.05, 0.05, 0.50)\textrm{, and} \nonumber \\
E&=&(.05, 0, 0, 0, 0, 0, 0, 0.95) \nonumber.
\end{eqnarray}

We considered the lane size equal to $10.000$ sites. The measurements were averaged over $10$ random initial conditions over $10^6$ time-steps after a transient of $10^{5}$ time steps. We present the results in figure \ref{HeteroNaSch}.  

We propose a local imitation process s follows. Every $10^2$ time-steps $5\%$ is randomly chosen. During the following $10^2$ time-steps we record the velocity of this focal particle, $\bar{v}_f$, and the velocity of the particle in front of it, $\bar{v}_{t}$. If the target particle spends more than $80\%$ of the time inside its interaction radius, considered as equal to the maximum velocity of the focal particle, then this particle imitates the parameters of its target with probability $p_t=\bar{v}_t/(\bar{v}_f+\bar{v}_t)$. Notice that $\tau_i$ here is $\tau_i \approx 2*10^2*(1/0.05)=4*10^3$ time-steps.

We present the evolutionary trajectory of the population in the parameters space in figure \ref{ImitationFixedPoint}. There is an attraction to the point $(2,0.18)$, which is the set of parameters related to the slower strategy in all heterogeneous populations.

To add mutation we randomly choose $5\%$ of the population to be subjected to the process in variable time-scales. With equal probability, we choose between $v_{max}$ or $p$ to be changed, with the former being subjected to a change of $\delta = 1$ and the latter to $\delta = 0.01$. The variation $\delta$ can be negative or positive with equal probability up to the bound of the parameters. The maximum velocity is defined in the set $v_{max}\in[1,\infty)$ and the randomization, $p\in[0,.99]$. Since the imitation dynamics happens each $10^3$ time-steps sampling the same portion of the population, the simulations were made over three different combinations of time-scales: after every $10^3$ imitation time-steps ($2*10^5$ time-steps), or $\tau_i/\tau_m = 0.001$; $10^3$ each imitation time-steps, or $\tau_i/\tau_m = 1000$, and; every time the mutation happens the imitation follows $\tau_i/\tau_m = 1$. We show the results of the evolution under these rules over $10^7$ time-steps in figure \ref{MutationFixedPoint}. 

\section{Example II: Multi-behavioral models}\label{ex2}

\begin{figure}
    \centering
    \includegraphics[scale=0.34]{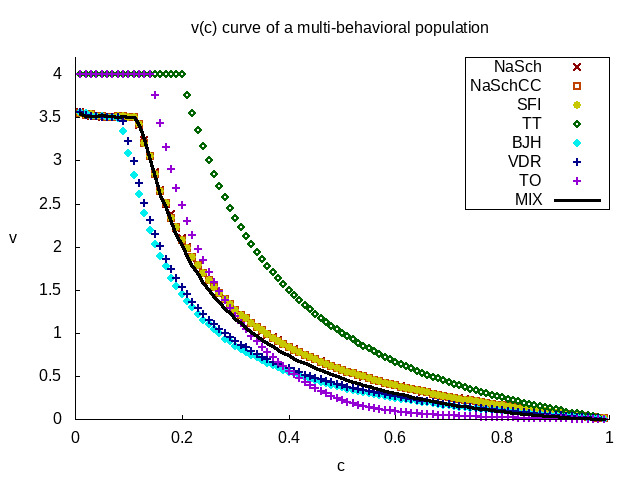}
    \includegraphics[scale=0.34]{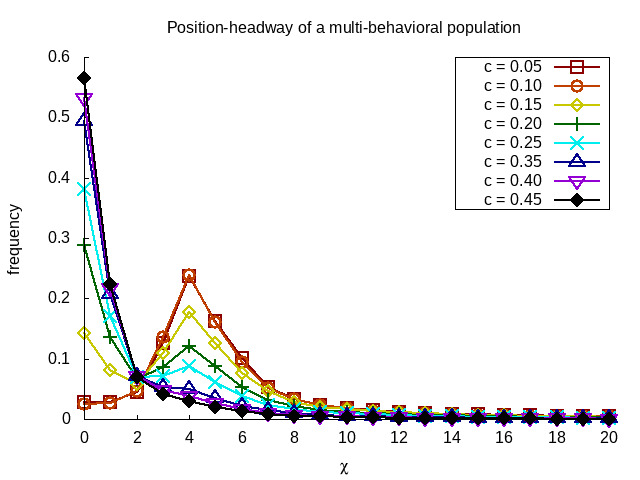}
    \caption{Simulational study of multi-behavioral populations. The average velocity in the top panel. The points represent heterogeneous populations of the algorithms cited in the main text, and the solid black line is a multi-behavioral population composed of equal amounts of all strategies. There is a change of behavior $c=0.4$, where the TO algorithm becomes slower than BJH model. Such change is unnoticed in the multi-behavioral population curve, but it indicates that the 'slower' strategy depends on the values of the parameters (including $\beta$) \textbf{and} the environmental conditions such as the density. Below, the relative distances histogram measure over many concentration values where two peaks struggle for dominance as the density increases. That is evidence of a multiplicity of dominating behaviors. The convergence of all curves near $\chi=2$ is interesting and we do not, at the present, have an explanation.}
    \label{MIX1}
\end{figure}

Despite the effort, incorporating more features into known models does not exhaust all the intrinsic characteristics of the many driving preferences. Trucks have lower accelerations than sports vehicles and decelerate upon comparably longer gaps. As the NaSch model does not incorporate parametrized acceleration, one needs a new algorithm to simulate trucks together with sports vehicles. With these considerations in mind, we define multi-behavioral populations a set of heterogeneous models emerging from different from algorithms. 

Consider seven, environmentally compatible,  CA traffic models taken from \cite{maerivoet2004non}. The already discussed NaSch model \cite{nagel1992cellular} which requires the parameters $v_{max}$ and $p$. A variation of the NaSch model that take into consideration two possible values for the randomization probability, the maximum velocity randomization, $p_f$, and otherwise $p$ (In the original model $p_f=0$), which is the cruise-control NaSch model (NaSch-CC) \cite{nagel1995emergent}. The stochastic model proposed by Fukui and Ishibashi (SFI)\cite{fukui1996traffic} make a similar modification to the NaSch model as the NaSch-CC model but the maximum velocity randomization is higher: $p_f>p$. (in the original paper $p=0$). Takayasu et al. proposed a model (TT) similar to the TASEP model but with a delayed acceleration in case the gap among vehicles is lower than a limit, $\chi_{s}$ \cite{takayasu19931}. This model requires two parameters $v_{max}$ and $\chi_{s}$ and in the original  work $v_{max}=1$ and $\chi_{s} = 2$. We will also consider the model proposed by Benjamim et al. (BJH)\cite{benjamin1996cellular} that implements a slow-to-start rule of temporal nature. A stopped vehicle will accelerate with probability $1-p_s$ and, in case it did not move, it will try again in the next time-step with probability $p_s$. The model proposed by Barlović et al.  (VDR)\cite{barlovic1998metastable,barlovic2003traffic} generalizes the NaSch model for an intuitive slow-to-start rule using the randomization of the stopped particle, $p_0$. Finally, the model proposed by Brilon e Wu (TO) \cite{brilon1999evaluation} uses temporal headways to obtain a more realistic particle-particle interaction compared to the NaSch model. It requires a safe time-gap $G_t$, the maximum velocity,  $v_{max}$, the randomization, $p$, an acceleration probability, $p_{acc}$, and a deceleration probability, $p_{dcc}$. Consult the original works for a thorough presentation of the models.
\begin{figure}
    \centering
    \includegraphics[scale=0.5]{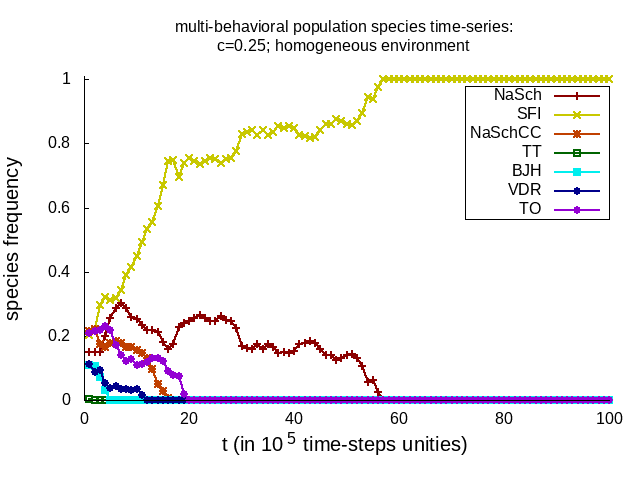}
    \includegraphics[scale=0.5]{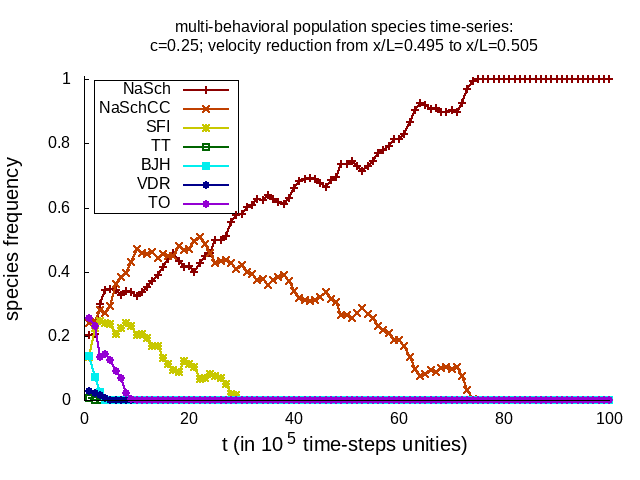}
    \caption{Time series of the frequency of the models in multi-behavioral populations under imitation. The global density is fixed in $c=0.25$. At the top, the environment is homogeneous. The winner is the SFI model. At the bottom, the environment has a reducing velocity that causes big variations in the local density. Surprisingly, the winner is the NaSch model.}
    \label{MIX2}
\end{figure}

One may join the models NaSch, NaSch-CC, SFI, and VDR in a single, comprehensive model requiring the set of parameters $(v_{max}, p, p_0, p_f)$  (for example, we define the VDR model in the subspace $(v_{max}, p, p_0, 0)$). The incorporation of the other models is not possible because they have different implementations. Nevertheless, one may join all algorithms introducing a discreet parameter $\beta$ that controls the behavior the particle will manifest expanding the space of parameters, $\Pi$, to contain all the parameters plus $\beta$: $\Pi_e = (v_{max}, p, p_s, p_0, p_f, p_{acc}, p_{dec}, \chi_s, \chi_t, \beta)$. For example, if $\beta = 1$ then we say the particle will follow the NaSch model $(v_{max},p)$. The reason to introduce this parameter is to introduce multi-behavioral populations as \textit{quenched disorder} in the same way as heterogeneous populations but in the extended parameters space, $\Pi_e$. Here, $\beta \in [1,7]$ with each integer associated with the algorithm of each model in the respective order.

The parameters were uniformly distributed in the intervals: $v_{max} \in [4,9]$; $p \in [0.1,0.5]$; $p_f \in [0,0.5]$; $p_s \in [0.9, 0.95]$; $p_0 \in [0.4, 0.7]$; $p_{acc} \in [0.9, 0.95]$; $p_{dcc} \in [0.8, 0.95]$; $\chi_s \in[0,2]$; $ \chi_{t} \in [1.2,1.3]$, and; $\beta \in [1,7]$. We present the velocity and the distance between particles curves in figure \ref{MIX1}. Remarkably, we have two distance maximums at $\chi=4$ and $\chi=0$,  exchange particles from one maximum to another with increasing density and the curves in all densities crossing a single point. Notice that at $c<0.4$, the BJH heterogeneous model is slower, but at $c>0.4$, the TO model becomes slower. These results suggest that multi-behavioral populations present multiple average velocity minimums that control the collective behavior at different conditions. 

As the extended space of parameter is $10$-dimensional we present a visualization of the evolution of this system with time series of the frequency of behaviors in figure \ref{MIX2}. As the handling of so many parameters is computationally expensive, we modify the environment to reduce the maximum velocity to $v_{max}=2$ in  $0.01\%$ of the road to simulate high-density bubbles. It was expected to observe the domination of two strategies in the two set-ups, but not the NaSch model as it is \textbf{not} one of the slower strategies at \textbf{any} concentrations (see figure \ref{MIX1}). This suggests that the evolution of multi-behavioral populations is a complex phenomenon that depends on the environmental conditions and the competing strategies. The evolution in these populations seems to have similar dynamics to the famous Axelrod's Tournament \cite{axelrod1980effective}.

 As the \textbf{Main result 2} foresee an orienting tendency at $\tau_i/\tau_m \rightarrow 0$ and the converse at $\tau_i/\tau_m \rightarrow \infty$, in our final experiment we define a ``order parameter '' as:
 
\begin{equation}\label{orderpaprameter}
    \theta=\frac{1}{N}\sum_{j=1}^{N}\frac{\textbf{m}_{j}\cdot \bar{\textbf{m}}}{|\textbf{m}_j||\bar{\textbf{m}}|}.
\end{equation}

$\textbf{m}_j$ is a vector characterized by the parameters of the $j$-th particle, $(v_{max}, p, p_s, p_0, p_f, p_{acc}, p_{dec}, \chi_s, \chi_t)$ but depends on the parameter $\beta$.

We consider $\beta$ as the parameter of a projection in $\Pi_e$. The NaSch model, for example, is represented as $m_i=(v_{max,i}, p_i, 0, 0, 0, 0, 0, 0, 0,1)$. This way, only variables subjected to evolutionary pressures enter in the calculation of the vector averaged over the population,  $\bar{\textbf{m}}$, which corresponds to partition $\Pi_e$ in dominions. Notice that the higher possible value of $\theta$ is one, which happens only in homogeneous population which corresponds in $\Pi_e$ to all particles to belong in a single ordered dominion.

\begin{figure*}[t]
    \centering
    \includegraphics[scale=0.33]{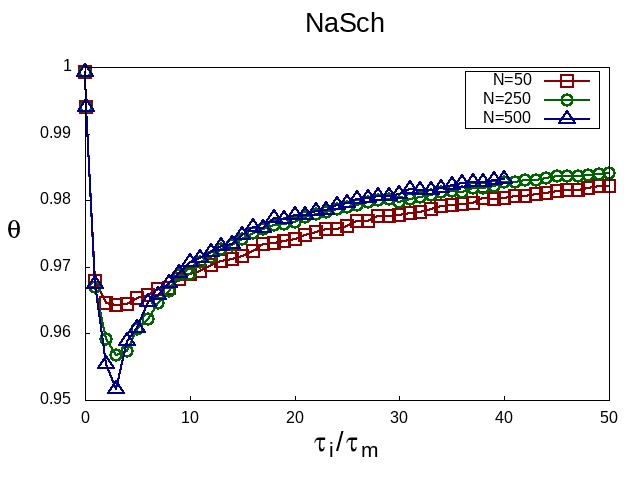}
    \includegraphics[scale=0.33]{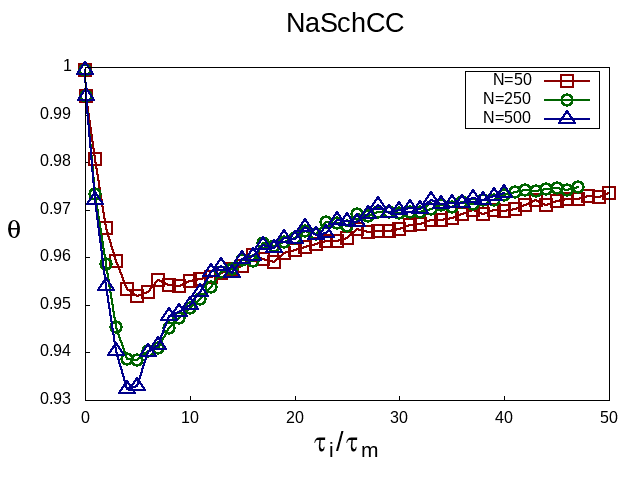}
    \includegraphics[scale=0.33]{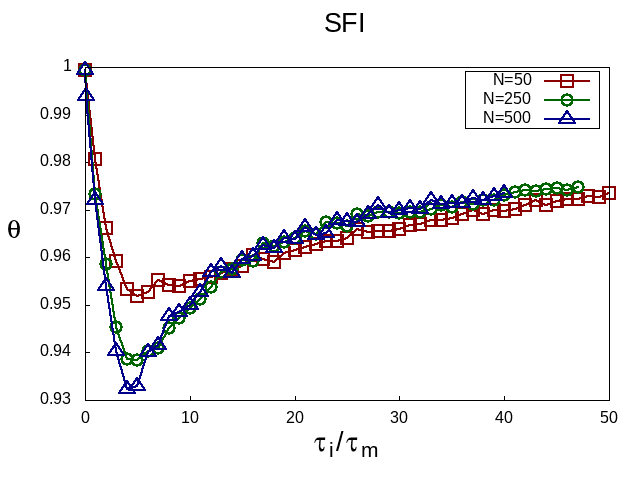}
    
    \includegraphics[scale=0.33]{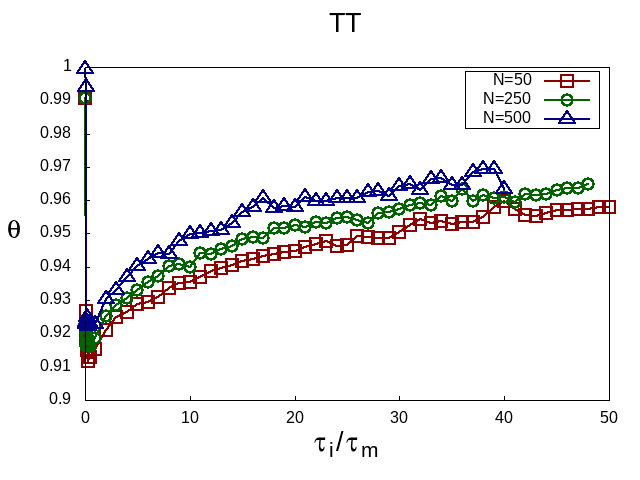}
    \includegraphics[scale=0.33]{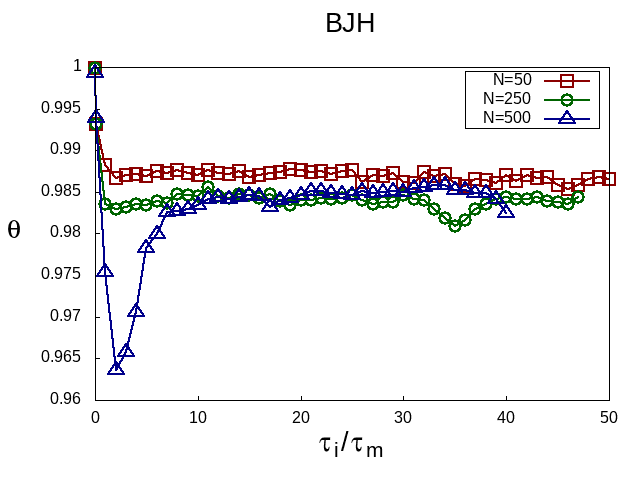}
    \includegraphics[scale=0.33]{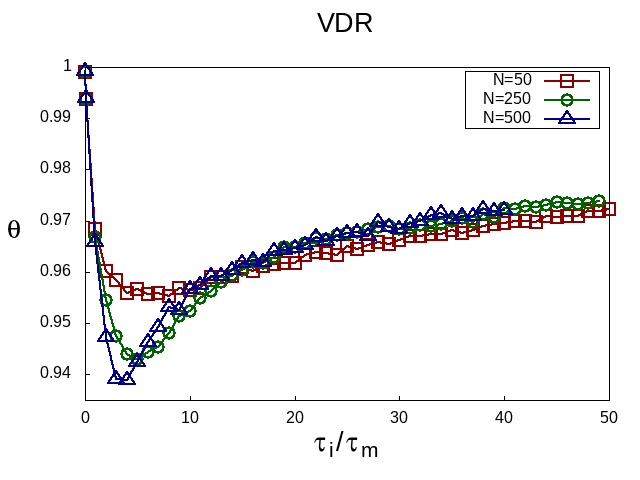}

    \includegraphics[scale=0.33]{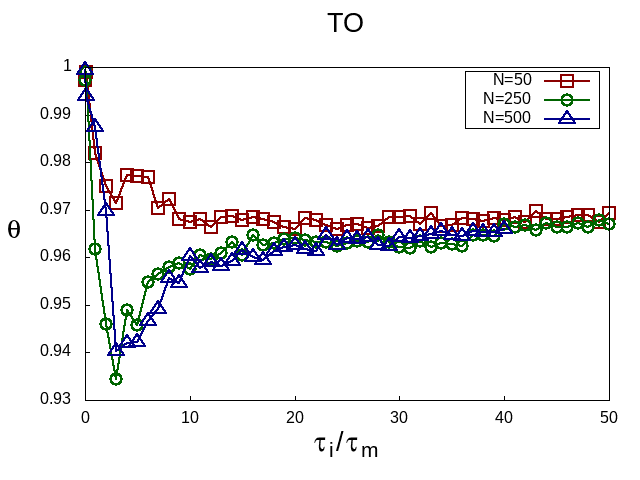}
    \includegraphics[scale=0.33]{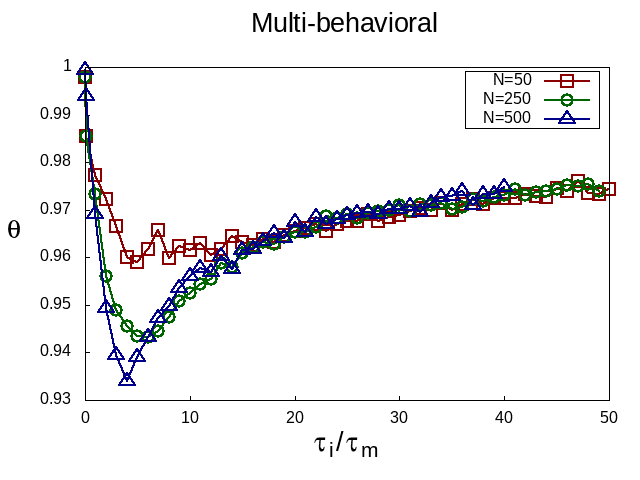}
    
    \caption{The parameter $\theta$ in function of $\tau_i/\tau_m$.  We updated the fraction of the population selected in each mutation process to $10\%$, and the imitation happens for all particles after $100$ steps. The species only change by imitation. The parameters and species initialization are the same as figure \ref{MIX2}. There is a transient of $10^{6}$ time-steps to begin the evolutionary processes, and after $10^{6}$ time-steps, we average $\theta$ over $10^{5}$ time-steps and over $50$ configurations. The similarity of all these curves is a consequence of the generality of the main results.}
    \label{OP}
\end{figure*}

The generality of the \textbf{Main result 2} implies that in very long runs, $\theta=\theta(\tau_{i}/\tau_m)$ and, additionally, in very large systems two asymptotic behaviors are expected: $\theta(\tau_{i}/\tau_m\rightarrow 0) \rightarrow 1$ and $\theta(\tau_{i}/\tau_m\rightarrow \infty) \rightarrow \alpha(\beta)<1$. In the first case, the imitation dynamics drive the system to a global orientation of vectors and species. In the second case, the mutation dynamics dominates and disorient the vectors. The precise value in this limit depends on the scope of the parameters of the models initially present in the simulation. We illustrate the curve $\theta(\tau_{i}/\tau_m)$ for different sizes of the system  at a concentration fixed at $c=0.05$ in figure \ref{OP}.

The results suggest new phenomena in the neighborhood $\tau_i/\tau_m \in [0.01,10]$ where a fall for very small $\theta$ gets more pronounced as we increase the size of the system. Notice that the smaller value $\theta$ can get is not related to a random disposition of the vectors, but the population dividing itself in ordered dominions with different directions and this is responsible for the fall at low $\tau_i/\tau_m$. Even after a long time there is coexistence of behaviors. By similarity to the \textit{n-vector} model, there is no reason to expect a ``real'' phase transition here, but further study on the stability of these coexistence and the universal shape of the curves should be addressed in future works.  

\section{Discussions and conclusion}\label{diss}

Single-laned heterogeneous traffic models under periodic boundary conditions satisfying properties \textbf{P1}, \textbf{P2} and \textbf{P3} exists a stable, steady-state at low densities. In this state, the average velocity of all vehicles is equal to the average velocity of the slower particle in the system and the existence of an average distance headway between particles which details depend on the algorithm. We use this to show a link between the net of interactions among the particles in the system and a directed, random graph with a single leader when the density of particles is low and a directed, cyclical, random graph when the density is high.

The introduction of local imitation dynamics as a mechanism to change the parameters of the particles to couple with the parameters of a particle it is interacting with promotes the evolution of behavior in the system transforming the quenched disorder that introduced heterogeneity in the system into an annealed disorder driven towards the most successful set of parameters. Surprisingly, the most successful strategy in the population is the slower strategy. Therefore, at low densities, the final outcome of the evolution of behaviors by ``social contact'' in these systems is a homogeneous state formed with the slower strategy initially in the population, regardless of the algorithm of imitation or the nature of the subtended social pressures, given that this process is local and random drift is allowed. At high densities, this result also holds, but the time needed to reach it can be very big as the process acquire similar characteristics to a $M$-dimensional random walk, where $M$ is the number of different behaviors.  

The mutation promotes independent, random changes in the parameters of the particles. The imitation dynamics brings into discussion the ``self-improvement'' towards shorter commutes and the mutation dynamics consider changes due to random factors not associated with the commute time minimization.  Associating to each particle, requiring $m$ parameters,  a $m$-dimensional vector defined in the space of parameters, we make a physical analogy: the imitation dynamics align vectors, and the mutation dynamics misalign these vectors. This suggest a similarity to the statistical mechanical \textit{n-vector} model with $n=m$. If imitation and mutation occur at characteristic times $\tau_i$ and $\tau_m$, respectively, then at low densities the asymptotic behavior of this coupling a is ordering of the vectors at $\tau_i/\tau_m \rightarrow 0$ and orientational disorder at $\tau_i/\tau_m \rightarrow \infty$. What happens in-between deserves further study.

The similarity is not rigorous as the interactions are not necessarily first-neighbors (but $R/L=0$ in the thermodynamic limit), are directed, the size of the interacting system is not constant, and we did not write down a \textit{Hamiltonian operator} for the system. This concern is important at low densities where the mutation process effects are \textbf{not} trivially related to thermal fluctuations in spin chains.

The uni-dimensional character of the system simplifies the problem considerably. In higher dimensions (multi-lane models), the association to a graph theory is impossible with the tools we have used here because the order of particles is not dynamically conserved but in situations where the multidimensional character of the system does not manifest. More study is necessary to include these systems.

About uni-dimensional continuous models, the heuristics used in the arguments are valid in continuous models and, therefore, the main results should also be valid in these systems. Nevertheless, the properties \textbf{P1}, \textbf{P2}, and \textbf{P3} are translated in continuous language as the following. Suppose a piece of the system, $\rho(x(t),t)$, evolves in time according to an operator $\mathcal{H}_t$ such that $\rho(x(t),t)$ can be approximated by an application of $\mathcal{H}$ in some state $\rho(x(0),0)$: $\rho(x(t),t)=\mathcal{H}_t\rho(x(0),0)$.  The continuous version of \textbf{P1} is the explicitly dependency of $\mathcal{H}_t$ as $\mathcal{H}_t=\mathcal{H}_t(y_1,y_2...y_m,\alpha_1,\alpha_2,...)$ where $y_i$ is the value of the $i$-th parameter of the model ($1\le  i \le m$) and the  $\alpha$'s are parameters related to the state of the system in front of $x(t)$ (such as local density and average velocity) at all times. Let $x_i(0)$ and $x_j(0)$ the locations of  different pieces of the system in $t=0$ such that $x_i(0) > x_j(0)$, then the continuous version of property \textbf{P2} is simply the non-holonomic constraint $x_i(t) > x_j(t)$ for all $t$. Property \textbf{P3} do not need translation to the continuous language but the individual to the local density of particles. To the best of our knowledge, no continuous models explicitly present these conditions.

From an empirical perspective, one may argue that automated vehicles would improve traffic as they adapt their strategy to reduce the commute time safely. Our results shed skepticism on the viability of this solution because, as long as there are non-automated vehicles (slow strategies) on the road, the automatization does not make significant changes in the traffic flow. 

Concerning the evolution of behaviors, one may argue that parameters such as $v_{max}$ have a physical interpretation of the externally given limit that all particles in the system are supposed to be able to reach, so it makes little sense to let it ``evolve''.

We observe that all mathematical and computational models use experimental parameters of different natures to calibrate their results. Some are directly related to the driver's preferences such as the parameter $p$, which can ``evolve''.  The study of the evolution of multi-behavioral models is important to understand phenomena related to the mixing of different populations. 

\section{Acknowledgments}
We are thankful to CNPq, CAPES, and FAPEMIG for the financial support.

\bibliographystyle{unsrtnat}
\bibliography{references.bib}

\begin{thebibliography}{62}
\providecommand{\natexlab}[1]{#1}
\providecommand{\url}[1]{\texttt{#1}}
\expandafter\ifx\csname urlstyle\endcsname\relax
  \providecommand{\doi}[1]{doi: #1}\else
  \providecommand{\doi}{doi: \begingroup \urlstyle{rm}\Url}\fi

\bibitem[Gino et~al.(2017)Gino, Staats, Jachimowicz, Lee, and
  Menges]{longcommute}
F.~Gino, B.R. Staats, J.~Jachimowicz, J.~Lee, and J.~Menges.
\newblock Reclaim your commute.
\newblock \emph{Harvard Bus. Rev.}, 95\penalty0 (5):\penalty0 149--153, 2017.

\bibitem[Peters et~al.(2004)Peters, Von~Klot, Heier, Trentinaglia, H{\"o}rmann,
  Wichmann, and L{\"o}wel]{peters2004exposure}
Annette Peters, Stephanie Von~Klot, Margit Heier, Ines Trentinaglia, Allmut
  H{\"o}rmann, H~Erich Wichmann, and Hannelore L{\"o}wel.
\newblock Exposure to traffic and the onset of myocardial infarction.
\newblock \emph{New England Journal of Medicine}, 351\penalty0 (17):\penalty0
  1721--1730, 2004.

\bibitem[Weisbrod et~al.(2003)Weisbrod, Vary, and Treyz]{weisbrod2003measuring}
Glen Weisbrod, Don Vary, and George Treyz.
\newblock Measuring economic costs of urban traffic congestion to business.
\newblock \emph{Transportation research record}, 1839\penalty0 (1):\penalty0
  98--106, 2003.

\bibitem[Lighthill and Whitham(1955{\natexlab{a}})]{lighthill1955kinematicI}
Michael~James Lighthill and G~Be Whitham.
\newblock On kinematic waves i. flood movement in long rivers.
\newblock \emph{Proceedings of the Royal Society of London. Series A.
  Mathematical and Physical Sciences}, 229\penalty0 (1178):\penalty0 281--316,
  1955{\natexlab{a}}.

\bibitem[Lighthill and Whitham(1955{\natexlab{b}})]{lighthill1955kinematic}
Michael~James Lighthill and Gerald~Beresford Whitham.
\newblock On kinematic waves ii. a theory of traffic flow on long crowded
  roads.
\newblock \emph{P. Roy. Soc. Lond. A Mat.}, 229\penalty0 (1178):\penalty0
  317--345, 1955{\natexlab{b}}.

\bibitem[Kerner(2012)]{kerner2012physics}
Boris~S Kerner.
\newblock \emph{The physics of traffic: empirical freeway pattern features,
  engineering applications, and theory}.
\newblock Springer, 2012.

\bibitem[Piccoli and Tosin(2009)]{piccoli2009vehicular}
Benedetto Piccoli and Andrea Tosin.
\newblock Vehicular traffic: A review of continuum mathematical models.
\newblock \emph{Encyclopedia of Complexity and Systems Science}, 22:\penalty0
  9727--9749, 2009.

\bibitem[Nagatani(2002)]{nagatani2002physics}
Takashi Nagatani.
\newblock The physics of traffic jams.
\newblock \emph{Rep. Prog. Phys.}, 65\penalty0 (9):\penalty0 1331, 2002.

\bibitem[Chowdhury et~al.(2000)Chowdhury, Santen, and
  Schadschneider]{chowdhury2000statistical}
Debashish Chowdhury, Ludger Santen, and Andreas Schadschneider.
\newblock Statistical physics of vehicular traffic and some related systems.
\newblock \emph{Phys. Rep.}, 329\penalty0 (4-6):\penalty0 199--329, 2000.

\bibitem[Kerner and Rehborn(1996)]{kerner1996experimental}
Boris~S Kerner and Hubert Rehborn.
\newblock Experimental features and characteristics of traffic jams.
\newblock \emph{Phys. Rev. E}, 53\penalty0 (2):\penalty0 R1297, 1996.

\bibitem[Maerivoet and De~Moor(2005)]{maerivoet2005cellular}
Sven Maerivoet and Bart De~Moor.
\newblock Cellular automata models of road traffic.
\newblock \emph{Physics reports}, 419\penalty0 (1):\penalty0 1--64, 2005.

\bibitem[Nagel(1996)]{nagel1996particle}
Kai Nagel.
\newblock Particle hopping models and traffic flow theory.
\newblock \emph{Physical review E}, 53\penalty0 (5):\penalty0 4655, 1996.

\bibitem[Hoogendoorn and Bovy(2001)]{hoogendoorn2001state}
Serge~P Hoogendoorn and Piet~HL Bovy.
\newblock State-of-the-art of vehicular traffic flow modelling.
\newblock \emph{Proceedings of the Institution of Mechanical Engineers, Part I:
  Journal of Systems and Control Engineering}, 215\penalty0 (4):\penalty0
  283--303, 2001.

\bibitem[Brackstone and McDonald(1999)]{brackstone1999car}
Mark Brackstone and Mike McDonald.
\newblock Car-following: a historical review.
\newblock \emph{Transportation Research Part F: Traffic Psychology and
  Behaviour}, 2\penalty0 (4):\penalty0 181--196, 1999.

\bibitem[Prigogine and Herman(1971)]{prigogine1971kinetic}
Ilya Prigogine and Robert Herman.
\newblock Kinetic theory of vehicular traffic.
\newblock Technical report, 1971.

\bibitem[Maerivoet and De~Moor(2004)]{maerivoet2004non}
Sven Maerivoet and Bart De~Moor.
\newblock Non-concave fundamental diagrams and phase transitions in a
  stochastic traffic cellular automaton.
\newblock \emph{The European Physical Journal B-Condensed Matter and Complex
  Systems}, 42\penalty0 (1):\penalty0 131--140, 2004.

\bibitem[Wolfram(1983)]{wolfram1983statistical}
Stephen Wolfram.
\newblock Statistical mechanics of cellular automata.
\newblock \emph{Rev. Mod. Phys.}, 55\penalty0 (3):\penalty0 601, 1983.

\bibitem[Papageorgiou(1998)]{papageorgiou1998some}
Markos Papageorgiou.
\newblock Some remarks on macroscopic traffic flow modelling.
\newblock \emph{Transportation Research Part A: Policy and Practice},
  32\penalty0 (5):\penalty0 323--329, 1998.

\bibitem[Ben-Naim et~al.(1994)Ben-Naim, Krapivsky, and Redner]{ben1994kinetics}
Eli Ben-Naim, Pavel~L Krapivsky, and Sidney Redner.
\newblock Kinetics of clustering in traffic flows.
\newblock \emph{Physical Review E}, 50\penalty0 (2):\penalty0 822, 1994.

\bibitem[Krug and Ferrari(1996)]{krug1996phase}
Joachim Krug and Pablo~A Ferrari.
\newblock Phase transitions in driven diffusive systems with random rates.
\newblock \emph{Journal of Physics A: Mathematical and General}, 29\penalty0
  (18):\penalty0 L465, 1996.

\bibitem[Helbing and Tilch(2008)]{helbing2008power}
Dirk Helbing and Benno Tilch.
\newblock A power law for the duration of high-flow states in heterogeneous
  traffic flows.
\newblock \emph{arXiv preprint arXiv:0807.3710}, 2008.

\bibitem[Krug(2000)]{krug2000phase}
Joachim Krug.
\newblock Phase separation in disordered exclusion models.
\newblock \emph{Brazilian Journal of Physics}, 30\penalty0 (1):\penalty0
  97--104, 2000.

\bibitem[Barma(2006)]{barma2006driven}
Mustansir Barma.
\newblock Driven diffusive systems with disorder.
\newblock \emph{Physica A: Statistical Mechanics and its Applications},
  372\penalty0 (1):\penalty0 22--33, 2006.

\bibitem[Ramana and Jabari(2020)]{ramana2020traffic}
A~Sai~Venkata Ramana and Saif~Eddin Jabari.
\newblock Traffic flow with multiple quenched disorders.
\newblock \emph{Physical Review E}, 101\penalty0 (5):\penalty0 052127, 2020.

\bibitem[Ramana and Jabari(2021)]{ramana2021power}
A~Sai~Venkata Ramana and Saif~Eddin Jabari.
\newblock Power laws and phase transitions in heterogenous car following with
  reaction times.
\newblock \emph{Physical Review E}, 103\penalty0 (3):\penalty0 032202, 2021.

\bibitem[Hagstrom and Abrams(2001)]{hagstrom2001characterizing}
Jane~N Hagstrom and Robert~A Abrams.
\newblock Characterizing braess's paradox for traffic networks.
\newblock In \emph{ITSC 2001. 2001 IEEE Intelligent Transportation Systems.
  Proceedings (Cat. No. 01TH8585)}, pages 836--841. IEEE, 2001.

\bibitem[Karlin and Peres(2017)]{karlin2017game}
Anna~R Karlin and Yuval Peres.
\newblock \emph{Game Theory, Alive}, volume 101.
\newblock American Mathematical Soc., 2017.

\bibitem[Iwamura and Tanimoto(2018)]{iwamura2018complex}
Yoshiro Iwamura and Jun Tanimoto.
\newblock Complex traffic flow that allows as well as hampers lane-changing
  intrinsically contains social-dilemma structures.
\newblock \emph{Journal of Statistical Mechanics: Theory and Experiment},
  2018\penalty0 (2):\penalty0 023408, 2018.

\bibitem[Tanimoto et~al.(2014)Tanimoto, Fujiki, Wang, Hagishima, and
  Ikegaya]{tanimoto2014dangerous}
Jun Tanimoto, Takuya Fujiki, Zhen Wang, Aya Hagishima, and Naoki Ikegaya.
\newblock Dangerous drivers foster social dilemma structures hidden behind a
  traffic flow with lane changes.
\newblock \emph{J. Stat. Mech-Theory E}, 2014\penalty0 (11):\penalty0 P11027,
  2014.

\bibitem[Tanimoto and An(2019)]{tanimoto2019improvement}
Jun Tanimoto and Xie An.
\newblock Improvement of traffic flux with introduction of a new lane-change
  protocol supported by intelligent traffic system.
\newblock \emph{Chaos, Solitons \& Fractals}, 122:\penalty0 1--5, 2019.

\bibitem[Tanimoto and Nakamura(2016)]{tanimoto2016social}
Jun Tanimoto and Kousuke Nakamura.
\newblock Social dilemma structure hidden behind traffic flow with route
  selection.
\newblock \emph{Physica A}, 459:\penalty0 92--99, 2016.

\bibitem[Nakata et~al.(2010)Nakata, Yamauchi, Tanimoto, and
  Hagishima]{nakata2010dilemma}
Makoto Nakata, Atsuo Yamauchi, Jun Tanimoto, and Aya Hagishima.
\newblock Dilemma game structure hidden in traffic flow at a bottleneck due to
  a 2 into 1 lane junction.
\newblock \emph{Physica A: Statistical Mechanics and its Applications},
  389\penalty0 (23):\penalty0 5353--5361, 2010.

\bibitem[Yamauchi et~al.(2009)Yamauchi, Tanimoto, Hagishima, and
  Sagara]{yamauchi2009dilemma}
Atsuo Yamauchi, Jun Tanimoto, Aya Hagishima, and Hiroki Sagara.
\newblock Dilemma game structure observed in traffic flow at a 2-to-1 lane
  junction.
\newblock \emph{Physical Review E}, 79\penalty0 (3):\penalty0 036104, 2009.

\bibitem[Sim{\~a}o and Wardil(2021)]{simao2021social}
Ricardo Sim{\~a}o and Lucas Wardil.
\newblock Social dilemma in traffic with heterogeneous drivers.
\newblock \emph{Physica A: Statistical Mechanics and its Applications},
  561:\penalty0 125235, 2021.

\bibitem[Sigmund and Nowak(1999)]{sigmund1999evolutionary}
Karl Sigmund and Martin~A Nowak.
\newblock Evolutionary game theory.
\newblock \emph{Current Biology}, 9\penalty0 (14):\penalty0 R503--R505, 1999.

\bibitem[Weibull(1997)]{weibull1997evolutionary}
J{\"o}rgen~W Weibull.
\newblock \emph{Evolutionary game theory}.
\newblock MIT press, 1997.

\bibitem[Hofbauer et~al.(1998)Hofbauer, Sigmund,
  et~al.]{hofbauer1998evolutionary}
Josef Hofbauer, Karl Sigmund, et~al.
\newblock \emph{Evolutionary games and population dynamics}.
\newblock Cambridge university press, 1998.

\bibitem[Vincent and Brown(2005)]{vincent2005evolutionary}
Thomas~L Vincent and Joel~S Brown.
\newblock \emph{Evolutionary game theory, natural selection, and Darwinian
  dynamics}.
\newblock Cambridge University Press, 2005.

\bibitem[Nowak(2006)]{nowak2006evolutionary}
Martin~A Nowak.
\newblock \emph{Evolutionary dynamics: exploring the equations of life}.
\newblock Harvard university press, 2006.

\bibitem[Perc et~al.(2013)Perc, G{\'o}mez-Gardenes, Szolnoki, Flor{\'\i}a, and
  Moreno]{perc2013evolutionary}
Matja{\v{z}} Perc, Jes{\'u}s G{\'o}mez-Gardenes, Attila Szolnoki, Luis~M
  Flor{\'\i}a, and Yamir Moreno.
\newblock Evolutionary dynamics of group interactions on structured
  populations: a review.
\newblock \emph{Journal of the royal society interface}, 10\penalty0
  (80):\penalty0 20120997, 2013.

\bibitem[Hofbauer and Sigmund(2003)]{hofbauer2003evolutionary}
Josef Hofbauer and Karl Sigmund.
\newblock Evolutionary game dynamics.
\newblock \emph{Bulletin of the American mathematical society}, 40\penalty0
  (4):\penalty0 479--519, 2003.

\bibitem[Ferreira(2002)]{ferreira2002mutation}
C{\^a}ndida Ferreira.
\newblock Mutation, transposition, and recombination: An analysis of the
  evolutionary dynamics.
\newblock In \emph{JCIS}, pages 614--617, 2002.

\bibitem[Foster and Young(1990)]{foster1990stochastic}
Dean Foster and Peyton Young.
\newblock Stochastic evolutionary game dynamics.
\newblock \emph{Theoretical population biology}, 38\penalty0 (2):\penalty0
  219--232, 1990.

\bibitem[Lieberman et~al.(2005)Lieberman, Hauert, and
  Nowak]{lieberman2005evolutionary}
Erez Lieberman, Christoph Hauert, and Martin~A Nowak.
\newblock Evolutionary dynamics on graphs.
\newblock \emph{Nature}, 433\penalty0 (7023):\penalty0 312--316, 2005.

\bibitem[Szab{\'o} and Fath(2007)]{szabo2007evolutionary}
Gy{\"o}rgy Szab{\'o} and Gabor Fath.
\newblock Evolutionary games on graphs.
\newblock \emph{Physics reports}, 446\penalty0 (4-6):\penalty0 97--216, 2007.

\bibitem[Shakarian et~al.(2012)Shakarian, Roos, and
  Johnson]{shakarian2012review}
Paulo Shakarian, Patrick Roos, and Anthony Johnson.
\newblock A review of evolutionary graph theory with applications to game
  theory.
\newblock \emph{Biosystems}, 107\penalty0 (2):\penalty0 66--80, 2012.

\bibitem[Fu et~al.(2009)Fu, Wang, Nowak, and Hauert]{fu2009evolutionary}
Feng Fu, Long Wang, Martin~A Nowak, and Christoph Hauert.
\newblock Evolutionary dynamics on graphs: Efficient method for weak selection.
\newblock \emph{Physical Review E}, 79\penalty0 (4):\penalty0 046707, 2009.

\bibitem[Nowak and Sigmund(2004)]{nowak2004evolutionary}
Martin~A Nowak and Karl Sigmund.
\newblock Evolutionary dynamics of biological games.
\newblock \emph{science}, 303\penalty0 (5659):\penalty0 793--799, 2004.

\bibitem[Vicsek and Zafeiris(2012)]{vicsek2012collective}
Tam{\'a}s Vicsek and Anna Zafeiris.
\newblock Collective motion.
\newblock \emph{Physics reports}, 517\penalty0 (3-4):\penalty0 71--140, 2012.

\bibitem[Gr{\'e}goire et~al.(2003)Gr{\'e}goire, Chat{\'e}, and
  Tu]{gregoire2003moving}
Guillaume Gr{\'e}goire, Hugues Chat{\'e}, and Yuhai Tu.
\newblock Moving and staying together without a leader.
\newblock \emph{Physica D: Nonlinear Phenomena}, 181\penalty0 (3-4):\penalty0
  157--170, 2003.

\bibitem[Doostmohammadi et~al.(2018)Doostmohammadi, Ign{\'e}s-Mullol, Yeomans,
  and Sagu{\'e}s]{doostmohammadi2018active}
Amin Doostmohammadi, Jordi Ign{\'e}s-Mullol, Julia~M Yeomans, and Francesc
  Sagu{\'e}s.
\newblock Active nematics.
\newblock \emph{Nature communications}, 9\penalty0 (1):\penalty0 1--13, 2018.

\bibitem[Gr{\'e}goire and Chat{\'e}(2004)]{gregoire2004onset}
Guillaume Gr{\'e}goire and Hugues Chat{\'e}.
\newblock Onset of collective and cohesive motion.
\newblock \emph{Physical review letters}, 92\penalty0 (2):\penalty0 025702,
  2004.

\bibitem[Nagel and Schreckenberg(1992)]{nagel1992cellular}
Kai Nagel and Michael Schreckenberg.
\newblock A cellular automaton model for freeway traffic.
\newblock \emph{J. Phys. I}, 2\penalty0 (12):\penalty0 2221--2229, 1992.

\bibitem[Evans(1997)]{evans1997exact}
MR~Evans.
\newblock Exact steady states of disordered hopping particle models with
  parallel and ordered sequential dynamics.
\newblock \emph{Journal of Physics A: Mathematical and General}, 30\penalty0
  (16):\penalty0 5669, 1997.

\bibitem[Nagel and Paczuski(1995)]{nagel1995emergent}
Kai Nagel and Maya Paczuski.
\newblock Emergent traffic jams.
\newblock \emph{Physical Review E}, 51\penalty0 (4):\penalty0 2909, 1995.

\bibitem[Fukui and Ishibashi(1996)]{fukui1996traffic}
Minoru Fukui and Yoshihiro Ishibashi.
\newblock Traffic flow in 1d cellular automaton model including cars moving
  with high speed.
\newblock \emph{Journal of the Physical Society of Japan}, 65\penalty0
  (6):\penalty0 1868--1870, 1996.

\bibitem[Takayasu and Takayasu(1993)]{takayasu19931}
Misako Takayasu and Hideki Takayasu.
\newblock 1/f noise in a traffic model.
\newblock \emph{Fractals}, 1\penalty0 (04):\penalty0 860--866, 1993.

\bibitem[Benjamin et~al.(1996)Benjamin, Johnson, and Hui]{benjamin1996cellular}
Simon~C Benjamin, Neil~F Johnson, and PM~Hui.
\newblock Cellular automata models of traffic flow along a highway containing a
  junction.
\newblock \emph{Journal of Physics A: Mathematical and General}, 29\penalty0
  (12):\penalty0 3119, 1996.

\bibitem[Barlovic et~al.(1998)Barlovic, Santen, Schadschneider, and
  Schreckenberg]{barlovic1998metastable}
Robert Barlovic, Ludger Santen, Andreas Schadschneider, and Michael
  Schreckenberg.
\newblock Metastable states in cellular automata for traffic flow.
\newblock \emph{Eur. Phys. J. B}, 5\penalty0 (3):\penalty0 793--800, 1998.

\bibitem[Barlovic(2003)]{barlovic2003traffic}
Robert Barlovic.
\newblock \emph{Traffic Jams: Cluster Formation in Low-Dimensional Cellular
  Automata Models for Highway and City Traffic}.
\newblock PhD thesis, 2003.

\bibitem[Brilon and Wu(1999)]{brilon1999evaluation}
W~Brilon and N~Wu.
\newblock Evaluation of cellular automata for traffic flow simulation on
  freeway and urban streets.
\newblock In \emph{Traffic and Mobility}, pages 163--180. Springer, 1999.

\bibitem[Axelrod(1980)]{axelrod1980effective}
Robert Axelrod.
\newblock Effective choice in the prisoner's dilemma.
\newblock \emph{Journal of conflict resolution}, 24\penalty0 (1):\penalty0
  3--25, 1980.

\end{thebibliography}

\section{Appendix}\label{Ap}

This appendix is dedicated to show the results presented in the main text. The arguments are strongly heuristic and we shall state the result followed by its argument. We left the commentaries and the definitions to the main text.  

\textbf{Auxiliary result 1:} In  uni-dimensional, heterogeneous, particle-flow systems satisfying the propositions P1, P2, and P3, there is a stable steady state characterized by $\chi_{j}^{*}$  and $v^*$, for all $ 1 \le j \le N $, at densities where a \textbf{leader} can be distinguished.

\begin{proof}[ Argument ]
Suppose that the single slow particle in the population is the $(j+1)$-th particle. In this case, $\chi_{j,j+1}$ surely decreases, due to P3 if $\chi_{j,j+1}(t) > R_j$  as the $j$-th particle have greater velocity than that of the $(j+1)$-th particle by hypothesis, and P1 that implies that the $(j+1)$-th particle will not have its own velocity affected by the behavior of the $j$-th particle. On the other hand, according to P2 it must be true that $\chi_{j,j+1}(t) \ge 0$ for all $t$. Writing the temporal average of the relative distance between those particles, $\bar{\chi}_{j,j+1}$, over $M$ measurements of duration $\Delta t=1$, we have:
\begin{equation}\label{SteadyStatePos}
\bar{\chi}_{j,j+1}=\frac{1}{M }\sum_{i=1}^{M}\chi_{j,j+1}(t_i) \rightarrow \chi^{*}_{j}, \textrm{ when } M  \rightarrow\infty .
\end{equation}

The convergence holds (in average, not in individual measurements) because, on one hand, $\chi_{j,j+1}$ is 'attracted' to the interval [$0,R_{j}$] for all $ 1 \le j \le N $ as the particles are dynamically independent for $\chi_{j,j+1}<R_j$. Now, the properties \textbf{P1}, \textbf{P2} and \textbf{P3} do not fixate the behavior of the $j$-th particle when it is interacting ($\chi_{j,j+1}(t) \le R_j$) with another particle so we will briefly show that these intrinsic rules does not matter except in pathological cases.

Suppose that prior from the measurements above, we have made other measurements of the $\chi_{j,j+1}(t)$ and made a histogram of the frequencies of these states. Whether the update algorithm is deterministic or stochastic the measurements presented in equation \ref{SteadyStatePos} should have the same results as the obtained from the constructed histogram when measured during a interval sufficiently big. In this case suppose that  $\chi_{j,j+1}(t)$ obeys a distribution $p(\chi_{j,j+1})$ to be measured in the interval $[0,R_{j}]$ for the distance among those particles to be in this interval and let $q(\chi_{j,j+1})$ be the distribution for the distance to have a value such that $\chi_{j,j+1}(t)>R_j$ we may write:

\begin{equation}\label{eq1}
\bar{\chi}_{j,j+1}=\frac{1}{M }\sum_{i=1}^{M}\chi_{j,j+1}(t_i) = \sum_{i=d}^{R_{j}}ip(i)+\sum_{i=R_{j}+1}^{L}iq(i).
\end{equation}

But by \textbf{P3} and the dynamical independence when $\chi_{j,j+1}(t)>R_j$, $q(\chi_{j,j+1})$ must decrease as $\chi_{j,j+1}$ increases to values over $R_{j}$. In the thermodynamic limit, unless $q(\chi_{j,j+1})$ decreases at a rate slower than $i^{-2}$ we may either drop this term entirely, in which case the sum in \ref{eq1} will be a sum of integers limited in the interval $[0,R_{j}]$, or keep the few leading terms of the second summing in which case the sum in \ref{eq1} will be a sum of integers limited in the interval $[0,R_{j}+\theta]$ where it was taken the first $\theta$ terms above $R_{j}$. Either way, we have a distribution with finite variation measured over (arbitrarily big) $M$ independent events, so the average must exist, is unique for each particle and is defined as equal to $\chi^{*}_{j}$.

To argue that decreasing rates of the \textbf{two-particles distance distribution} slower than $i^{-2}$ are pathological, we observe that,  after the agglutination behind the slower particle by \textbf{P3} takes place, a polynomial behavior of the measurements may be observed when, in the interactions, there is a possible outcome that makes $\chi$ increase to higher values than $R$. The only way to achieve this obeying \textbf{P2} is stopping or reducing the velocity of the particle during a non-null time interval.  But if this interval is big enough that $\chi$ becomes higher than $R$, as required, then this would be a violation of \textbf{P3} that states that if the particles are dynamically independent than the velocities should be maximized. We reach a contradiction, so $q(\chi_{j,j+1})$ can be truncated in some finite, generally small, $\theta$. We observe that this type of behavior may simulate accidents occurrences in some model. So other then classify it as impossible we heather consider it as pathological here.

Directing our attention to the velocity of the $j$-th and $(j+1)$-th particles,  averaged over $M$ consecutive measurements of duration $\Delta t = 1$, we have that:

\begin{eqnarray}
\bar{v}_{j+1}M= \sum_{i=1}^{M} \Delta x_{j+1}=\sum_{i=1}^{M} x_{j+1}(t_{i+1})-x_{j+1}(t_i)& \\
=\sum_{i=1}^{M} (x_{j}(t_{i+1})-x_{j}(t_i))+\sum_{i=1}^{M}(\chi_{j}(t_{i+1}) - \chi_{j}(t_i)) \textrm{ \ \ \ \ }& \\
=\sum_{i=1}^{M}\Delta x_{j} - \sum_{i=1}^{M} \Delta \chi_{j,j+1} = \bar{v}_{j}M - \sum_{i=1}^{M} \Delta \chi_{j}\textrm{ \ \ \ \ \ \ \  }&.
\end{eqnarray}

We used the definition of measured averaged velocity and the identity $x_{k+1}(t)=x_{k}(t)+\chi_{k,k+1}(t)$. Because  $\sum_{i=1}^{M}\Delta \chi_{j,j+1}/M  \rightarrow 0$ as $M\rightarrow \infty$ by the definition of $\bar{\chi}_{j,j+1}$ ($\Delta \chi_{j,j+1} = \chi_{j}(t_{i+1}) - \chi_{j}(t_i)$ represent the difference between two random variables with the same average and variance) we reach the conclusion that the velocity of the fast particle is equal, on average, to the velocity of the leading slow particle:$\bar{v}_{j} = \bar{v}_{j+1}$.

Now we repeat the argument to all the particles behind these two until all have the same average velocity and have the relative distances fluctuating around the limit average values. To complete the argument we suppose that the fast particle initializes in front of the slow particle. As we are using periodic boundary conditions, all the fast particles in front of the slow one will loop and get behind the it and we are back in the previous situation. If there is more than one slow particle, we can repeat the proof starting at each slow particle. The indistinguishably among particles of the same strategy implies the same results for all and the outcome is a steady-state composed of clusters of particles led by the slow particles. 

We have required that densities low enough to distinguish a leader because we assumed that $E(\chi_{j,j+1})(t)=E(\chi_{j,j+1})(t+M)=\chi^{*}_{j}$ ($E(x)$ is the expected value of $x$) in the steady state for all $M$ positive. In other words, we have not considered the possibility of cycles or more complicated behavior that may be possible in densities where \textbf{all} particles may interact with each other.   
\end{proof}

\textbf{Auxiliary result 2:} In systems satisfying the conditions of the \textbf{Auxiliary result 1}, there is a graph, $\Omega$, that may represent the interaction nets on the system. $\Omega$ is a directed, random graph with a single leader.

 \begin{proof}[Argument]
 From \textbf{Auxiliary result 1} we know that there are an attracting steady-state where with the slowest strategy in the population as leaders. If this particle is the $j$-th, then the enumeration $\sigma=(j+1,j+2,...,N-1,N,1,...,j-1,j)$ is invariant under any dynamics of the particles that satisfies \textbf{P2} and represent the ordering of the particles with regard to the effective positions as one measures $x_{j}>x_{j-1}...>x_{j+1}$ up to a constant related to the beginning of the measurements. As $\chi_{i,k}(t)=x_{i}(t)-x_{k}(t)-d_i$, define the frequency of interactions between the $i$-th and the $k$-th particles ($x_{i}>x_{k}$), measured over an interval $T$, as $\omega_{i,k} = T_{\chi_{i,k}\le R_{k}}/T$, where $T_{\chi_{i,k}\le R_{k}}$ is the  the accumulated time in which the relative position is such that $\chi_{i,k}\le R_{k}$. 
 
 Now we construct $\Omega$ as a graph which takes the enumeration $\sigma$ as nodes and $\omega_{i,k}$ as the connectivity between the $i$ and $k$ nodes. $\Omega$ has a single leader because $\omega_{j,k} = 0$ for all $1 \le k \le N, k \ne j$, as the leader do not interact with any other particle.  It is random because $\omega_{i,k} = [0,1]$ for all $1 \le i,k \leq N, k \ne i$ and is directed by \textbf{P1} concluding the argument.
 \end{proof}

\textbf{Auxiliary result 3:} In systems satisfying the conditions of the \textbf{Auxiliary result 1}, the average velocity in the steady-state of any particle in a heterogeneous population is not higher than the average velocity of the slowest particle in this population.

\begin{proof}[Argument]
From \textbf{Auxiliary  result 1} all particles, in the steady-state, have the same average velocity and we can calculate this velocity for one of them. Again, from \textbf{Auxiliary  result 1} we know that the leader of the cluster is the slowest in the population, it follows that $v^{*}=\bar{v}_{s}$. 
\end{proof}
 
\textbf{Main result 1:}(Domination of the slowest) In systems under the conditions of the \textbf{Auxiliary result 1} under a local imitation process, the final outcome of this evolutionary process is a population using the strategy that returns slowest average velocity present in the initial population.

\begin{proof}[Argument] 

This result can be thought of as a direct consequence of the \textbf{Auxiliary result 2} in combination with the well-known result presented in \cite{fu2009evolutionary,lieberman2005evolutionary,nowak2004evolutionary}, that states that populations with individuals located in the nodes of directed graphs with single leaders undergoing local imitation processes are dominated by whatever strategy that occupy the leading position. As the probability of the slowest strategy to occupy the first position is one, this strategy will dominate the population. This view is somewhat unsatisfactory because the homogeneous population formed in the 'head' of the cluster locally invalidate the \textbf{Auxiliary result 1}.

Another interesting approach is as follows. The leader cannot imitate anyone as it interacts with no one. The next particle can only imitate the leader and, as the payoff measure is the average velocity of the individual, the \textbf{Auxiliary result 3} and the definition of the local imitation process guarantees that this change will eventually happen. After that, this particle will have no other strategy to imitate from, so its strategy becomes invariable under the imitation process. Repeating the argument to the particles behind them we conclude that the slowest strategy will erase the heterogeneous character of the system reproducing its own strategy. To complete the argument, we notice that as the strategies of the top particles get erased, it forms a patch of homogeneous slow population. If the concentration is such that, although a leader can be distinguished in the heterogeneous population, the top homogeneous patch can reach the bottom of the cluster the probability to choose a fast particle to become a slow particle is the same as the probability to choose a fast particle to become a slow particle as the selection is random. This means that the evolutionary dynamics now becomes a uni-dimensional random walk in which, for example, choosing a fast strategy to become slow corresponds to take a step to the right, and choosing a fast strategy to become a slow particle corresponds to take a step to the left. The details of the probabilities of going left or right in this random walk depends on the details of the imitation dynamics, but among the possible evolutionary fixed points where only one strategy remains, the states dominated by faster strategies are unreachable because the \textbf{Auxiliary result 1} implies that faster strategies do not lead stable structures. The only stable steady-state is, therefore, a state composed of only slow strategies. 
\end{proof}

\textbf{Main result 2:}(Domination of the slowest of all ) Systems under the conditions of the \textbf{Auxiliary result 1}, subjected to mutation associated to local imitation dynemics presents two distinct behaviors in the limits $\tau_i/\tau_m \rightarrow \infty$ and $\tau_i/\tau_m \rightarrow 0$.
In the former limit,  the long-time distribution will be analogous to a random walk. In the latter, the slowest strategies reachable in the parameter space will dominate the population.

\begin{proof}[Argument]

If $\tau_i/\tau_m \rightarrow \infty$, then the mutation process is more important than the imitation process which we neglect. The mutation process is a random walk in the space of parameters with no means to select the better fit by definition. This means that mutation alone would not provide the dynamics an effective \textit{drive} towards one direction or another. In case the parameters are contained in a closed space then eventually all the values will be visited and without the selection, the probability to measure any of then is equal to all. If the parameter, say $X_t$, is bounded in only one end, say $X_0$, define auxiliary variable $Y_t = X_t-X_0 $ if $X_0$ is a lower bound, or $Y_t = X_0-X_t $ if $X_0$ is an upper bound. With this transformation, the effect of the mutation dynamics in $Y_t$ is of a random walk that can only assume positive values. If neither end is bounded, then the result follows trivially from the definition of the process.

In the case $\tau_i/\tau_m \rightarrow 0$ and observing the independence of the imitation and the mutation processes, the \textbf{Main result 1} states that the imitation dynamics, that dominates the evolutionary dynamics at short intervals, will eventually bring the system to a homogeneous slow population. The mutation dynamics cause the appearance of new strategies that are perceived as faster, slower, or equal to the strategy dominating the population. Now, \textbf{Main result 1} states that the faster strategy will be washed out by the imitation dynamics regardless of how fast it is. On the other hand, if this particle mutates to a slower strategy, then the same result states that this strategy will dominate the population. Therefore, we have a stochastic process that \textit{drives} the system, in the strategies space, to the locations occupied by the parameters characteristics of the slower strategies, which do not need to be singled-valued. Notice that although the \textbf{Auxiliary result 2} is still valid, the ordering of the interaction's net is constantly reset by the mutation process, such that  $\Omega$ is no longer stationary.
\end{proof}

\end{document}